Structure of Matter and Quantum Chemistry

# Structures and Properties of *ß*-Titanium Doping Trace Transition Metal Elements: a Density Functional Theory Study

Jia Song[a,*], Luyu Wang[b], Liang Zhang[a], Kaiqi Wu[a],

Wenheng Wu[a]   and   Zhibin Gao[c,*]

[a]*Shanghai Engineering Research Center of 3D Printing Materials, Shanghai Research Institute of Materials, Shanghai 200437, China*

[b]*Institute of Fiber-Based New Energy Materials, The Key Laboratory of Advanced Textile Materials and Manufacturing Technology of Ministry of Education, College of Materials and Textiles, Zhengjia Sci-Tech University, Hangzhou 310018, China*

[c]*Department of Physics, National University of Singapore, Singapore 117551, Republic of Singapore*

*e-mail: skylve@t.shu.edu.cn; zhibin.gao@nus.edu.sg

**Abstract**–We systematically calculate the structure, formation enthalpy, formation free energy, elastic constants and electronic structure of $Ti_{0.98}X_{0.02}$ system by density functional theory (DFT) simulations to explore the effect of transition metal X (X= Ag, Cd, Co, Cr, Cu, Fe, Mn, Mo, Nb, Ni, Pd, Rh, Ru, Tc, and Zn) on the stability mechanism of *ß*-titanium. Based on our calculations, the results of formation enthalpy and free energy show that the adding trace X is beneficial to the thermodynamic stability of *ß*-titanium. This behavior is well explained by the density of state (DOS). However, the tetragonal shear moduli of $Ti_{0.98}X_{0.02}$ systems are negative, indicating that *ß*-titanium doping with a low concentration of X is still elastically unstable at 0 K. Therefore, we theoretically explain that *ß*-titanium doping with trace transition metal

X is unstable in the ground state.

**Introduction**

Due to titanium and its alloys possessing good strength-to-weight ratio, high heat resistance, and outstanding biocompatibility, they are widely used in numerous engineering fields, especially biomedicine applications [1–6]. However, on the one hand, traditional *α* type and *α+β* type titanium alloys have a higher elastic modulus, which can result in stress shielding effect. On the other hand, they also contain elements such as Al and V, which may take poisonous effect on biological cells [7,8]. Therefore, development of non-toxic and low-modulus *β*-titanium alloys has become a hot and attractive research topic in recent years.

In titanium alloys, generally different elements are added as alloying elements or are regarded as impurities. These elements have great effects on the properties of titanium alloys [3–11]. Many traditional experiments have been done to research the effect of alloying elements on the titanium alloys. Lee et al. [9] investigated rapid cooling texture of Ti-Nb alloys and surprisingly found that when the Nb content was less than 15 wt.%, the alloys were mainly composed of martensites *α′*; When the Nb content is in the range of 15~30 wt.%, there were oblique martensites *α″* in alloys; When the Nb content exceeds 30 wt.%, there were *β* phases in the alloys. Zhou et al. [10] studied the effect of element Ta on the properties of Ti alloys. Their results demonstrated that the elastic modulus curve of Ta-Ti alloys varied with Ta content and presented a "W" shape. When the Ta contents were 30 and 70 wt. %, the elastic modulus was the lowest, 69 and 70 GPa, respectively. Tang et al. [11] believed that the addition of element Zr could enhance the stability of the *β* phase, lower the martensite transformation temperature and suppress the generation of *ω* phase. In addition, they also found that in the Ti-Nb-Ta-Zr alloys, when the Nb/Ta ratio was 12 and the Zr content was 5 at.%, *ω* phase was completely suppressed and the alloys had a low elastic modulus.

With the continuous development of computer technology, the first-principles simulations based on quantum mechanics also get great progress. Using first principles to study titanium alloys has become hotter and hotter. Zou et al. [4] systematically investigated the bonding ability of solute atoms and Ti atom in the $Ti_{0.95}X_{0.05}$ systems (X=Al, Cr, V, Nb, and Mo) based on the DFT simulations. Their results suggested that the bonding strength of $Ti_{0.95}X_{0.05}$ increased in the order of Ti-Al<Ti-Cr<Ti-V<Ti-Nb<Ti-Mo. Zhou et al. [6] studied the phase stability of $Ti_{1-x}X_x$ alloys (X=substitutional Mo, Nb Al, Sn, Zr, Fe Co, and interstitial O) in *β* phase and *α* phase. Their results indicated that adding Mo, Nb, Fe, and Co increased the mechanical stabilize of *β*-titanium, whereas bcc Ti-Al, Ti-Sn, and Ti-Zr systems were mechanically unstable for all concentrations. For their *hcp* counterparts, Ti-Mo and Ti-Nb had lower mechanical than Ti-Al, Ti-O and Ti-Sn systems. Marker et al. [5] calculated elastic constants of five *bcc* Ti-X systems (X=Mo, Nb, Ta, Zr, and Sn) using first-principles simulations. Their calculations showed that according to the Born criteria, the structures of *β*-titanium were stabilized by adding 5.5 at.% Mo, 11.5 at.% Nb, 51.5 at.% Sn, 9.5 at.% Ta and 4.0 at.% Zn, respectively.

In this work, DFT simulations are performed to systematically investigate the *β*-titanium doping with trace transition metal elements X (X=Ag, Cd, Co, Cr, Cu, Fe, Mn, Mo, Nb, Ni, Pd, Rh, Ru, Tc, and Zn). The equilibrium volume, formation enthalpy and formation free energy are firstly calculated to study the effect of dopant X on the structure and thermodynamic properties of *β*-titanium. Secondly, the total density of states (TDOS) and partial density of states (PDOS) of $Ti_{0.98}X_{0.02}$ systems are estimated to explore the structural stability mechanism from the electronic level. Finally, the elastic constants of $Ti_{0.98}X_{0.02}$ systems are evaluated to research the effect of these transition elements on elastic stability of *β*-titanium.

## Methodology

### *Density functional theory calculations*

Vienna Ab-Inition Simulation Package (VASP) is used to conduct all DFT calculations [12–14]. The generalized gradient approximation (GGA) of Perdew–Burke–Ernzerhof (PBE) functional is used to calculate the exchange-correction energy of electron[15, 16]. The ion-electron interaction is described by the projector augmented wave (PAW) approach [17]. For consistency, the electronic wave functions are expanded in a plane-wave basis with a cutoff energy of 500 eV for all calculations, which is roughly 1.3 times higher than the default value [18]. The pseudopotentials describe the valence states of metal elements are as follows: Ti($3p^63d^24s^2$), Ag($4d^{10}5s^1$), Cd($4d^{10}5s^2$), Co($3d^74s^2$), Cr($3d^54s^1$), Cu($3d^{10}4s^1$), Fe($3d^64s^2$), Mn($3d^54s^2$), Mo($4s^24p^64d^55s^1$), Nb($4s^24p^64d^45s^1$), Ni($3d^84s^2$), Pd($4d^85p^2$), Rh($4d^75s^2$), Ru($4s^24p^64d^65s^2$), Tc($4s^24p^64d^55s^2$), and Zn($3d^{10}4s^2$), respectively. The Methfessel–Paxton method with smearing width of 0.05 is used to relax structure and atom positions of Ti$_{0.98}$X$_{0.02}$ system, while a tetrahedron method with Blöchl correction is used to get accurate energy and stress during the static calculations. To ensure a good precision in the calculation, the energy convergence criterion of the electronic self-consistency is set to $10^{-6}$ eV per atom, and the force convergence criterion is $10^{-2}$ eV/Å per atom. Spin-orbit coupling is included in all simulations, as implemented within the PAW method in the VASP code.

To research the effect of trace transition metal element on structure and stability of *β*-titanium, we construct a series of 54-atom body-centered cubic (bcc) supercells (based on 3×3×3 simple cubic cells) with one of the titanium atoms substituted by a metal element X atom (X=Ag, Cd, Co, Cr, Cu, Fe, Mn, Mo, Nb, Ni, Pd, Rh, Ru, Tc, and Zn). The atomic structure diagram of Ti$_{0.98}$X$_{0.02}$ system is shown in Fig. 1. The Brillouin zone is sampled by a 3×3×3 Monkhorst–Pack ***k***-point grid for Ti$_{0.98}$X$_{0.02}$ systems and by a 15×15×15 Monkhorst–Pack ***k***-point grid for pure Ti and X cell, respectively.

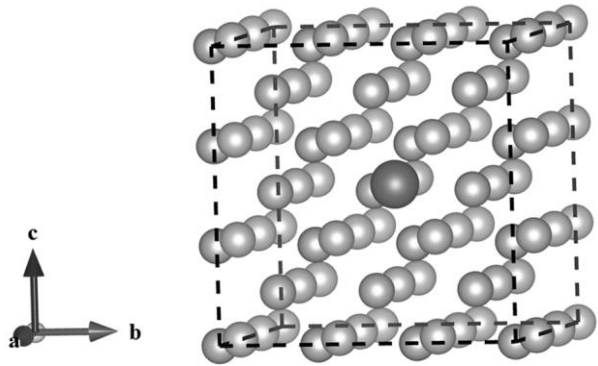

**Fig. 1.** The atomic structure diagram of Ti$_{0.98}$X$_{0.02}$. The dark grey ball represents the metal X atom and the rest light grey balls represent Ti atoms.

### *Stability of $Ti_{0.98}X_{0.02}$ systems*

As we well-known, the formation enthalpy can be used to preliminarily judge the stability of materials. The negative value of formation enthalpy means that $Ti_{0.98}X_{0.02}$ is thermodynamically stable, while a positive value indicates that it is unstable.[19] In this study, to make simulation results more reliable, the zero-point energy correction is added to formation enthalpy of $Ti_{0.98}X_{0.02}$ system, defined as follows [20,21],

$$\Delta H_f(Ti_{0.98}X_{0.02}) = E_{total}(Ti_{0.98}X_{0.02}) - 0.98E(Ti) - 0.02E(X) + \Delta ZPE, \quad (1)$$

in which, $E_{total}(Ti_{0.98}X_{0.02})$ is the total energy per atoms of $Ti_{0.98}X_{0.02}$ system, $E(Ti)$ and $E(X)$ are the energy per atom of Ti and transition metal X in their ground states: bcc Ti, fcc Ag, hcp Cd, hcp Co, bcc Cr, bcc Cu, bcc Fe, bcc Mn, bcc Mo, bcc Nb, fcc Ni, fcc Pd, fcc Rh, hcp Tc, and hcp Zn, $\Delta ZPE$ is the zero-point energy correction originated from lattice vibration.

Temperature is one of the important factors in the study of stability of materials, because most materials are used in a certain temperature range, not absolute zero. However, due to the limitation of calculation, this work roughly considers the influence of temperature by adding the contribution of configurational entropy within ideal mixing approximation, which is defined by [22],

$$S_{conf.}(x) = -k_B[x\ln(x) + (1-x)\ln(1-x)],$$

in which $x$ is the alloying element fraction of $Ti_{0.98}X_{0.02}$ system, 0.02, and $k_B$ is the Boltzmann constant. Therefore, the free energy of formation is estimated by,

$$\Delta F_f(Ti_{0.98}X_{0.02}) = \Delta H_f(Ti_{0.98}X_{0.02}) - TS_{conf.}(x)$$

Using the free energy of formation at a certain temperature, the temperature effect can be appropriately handled in the calculation of the phase stability of the alloy system. This method has been proven reasonable in previous report [6,22].

### *Elastic constants*

The elastic constants determine the stiffness of a crystal against an externally applied strain. In the case of small deformation, there is a quadratic dependence of the internal energy on the strain tensor. The elastic constants describe the quadratic relationship and are given by [23]

$$C_{ijkl} = \frac{1}{V}\left[\frac{\partial^2 E(V,\{\varepsilon_{mn}\})}{\partial \varepsilon_{ij} \partial \varepsilon_{kl}}\right]_{\varepsilon=0}, \quad (2)$$

in which, $E(V, \{\varepsilon_{mn}\})$ is the internal energy of the crystal after strain tensor $\varepsilon_{mn}$ is applied, $V$ is the volume of the unstrained crystal. Generally, the fourth-rank elastic constant $C_{ijkl}$ has no more than 21 independent components. The higher the symmetry of the crystal, the less the number of independent components.

For cubic $Ti_{0.98}X_{0.02}$ systems and $\beta$-titanium, there are three distinct, non-vanishing elastic constants, which are $C_{11}$, $C_{12}$ and $C_{44}$ [24,25]. This elastic tensor is determined by performing six finite distortions of the lattice and deriving the elastic constants from the strain-energy relationship. The applied strain modes are given in Table 1. The deformation magnitudes $\varepsilon$ from -0.016 to 0.016 in the step of 0.04 are used in the first and second strain modes, and $\varepsilon$ from -0.04 to 0.04 in the step of 0.01 are applied in the third strain mode [24,25].

**Table 1.** Parameterizations of the three strain modes used to calculate the three elastic constants of cubic $Ti_{0.98}X_{0.02}$ systems and $\beta$-titanium

| Strain | Parameters (unlisted $\varepsilon_i = 0$) | $\Delta E/V_0$ to $O(\gamma^2)$ |
|---|---|---|
| 1 | $\varepsilon_1=\varepsilon_2=\gamma$, $\varepsilon_3=(1+\gamma)^{-2}-1$ | $3(C_{11}-C_{12})\gamma^2$ |
| 2 | $\varepsilon_1=\varepsilon_2=\varepsilon_3=\gamma$ | $\frac{3}{2}(C_{11}+2C_{12})\gamma^2$ |
| 3 | $\varepsilon_6=2\gamma$, $\varepsilon_3=\gamma^2(4-\gamma^2)^{-1}$ | $2C_{44}\gamma^2$ |

Once the independent elastic constants of $Ti_{0.98}X_{0.02}$ and $\beta$-titanium are accurately calculated, the mechanical stability of these cubic structures can be determined by the following criterion [26]

$$C_{11} > 0, C_{44} > 0, C_{11} - C_{12} > 0, C_{11} + 2C_{12} > 0 \quad (3)$$

And the tetragonal shear modulus $C'$ of the crystal are also given by [6,27,28]

$$C' = (C_{11} - C_{22})/2 \tag{4}$$

Therefore, when the $C_{11}$, $C_{12}$, and $C_{44}$ are all greater than 0, the tetragonal shear modulus can be used to judge the stability of the cubic structure. A negative tetragonal shear modulus means that the cubic structure is unstable, and vice versa. And the lager the value of tetragonal shear modulus, the more stable the structure [6,27,28].

## Results and discussion

### *Structural optimization*

The equilibrium lattice parameter of *β*-titanium obtained from the zero-pressure geometry optimization is 3.252 Å, which is similar to experimental value of 3.283 Å with an absolute deviation of 0.9% [29], and also excellently consistent with previous literature value of 3.255 Å from DFT simulations [30]. Hence, we have reason to consider that the zero-pressure geometry optimization method in this study is rational. Using the same method, the equilibrium lattice parameters of $Ti_{0.98}X_{0.02}$ systems are obtained and summarized in Table 2. It can be found that *a*, *b* and *c* (lattice parameters of $Ti_{0.98}X_{0.02}$ systems) are almost equal, meaning that after *β*-titanium is doped with a low concentration of X, it still maintains bcc structure. What's more, it is obvious that different metal elements have different effects on the equilibrium volume of *β*-titanium. This phenomenon can be explained by the Wigner Seitz atomic radius. In this study, the Wigner Seitz radius are 1.323 Å for Ti, 1.503 Å for Ag, 1.577 Å for Cd, 1.302 Å for Co, 1.323 Å for Cr, 1.164 Å for Cu, 1.302 Å for Fe, 1.323 Å for Mn, 1.455 Å for Mo, 1.270 Å for Nb, 1.286 Å for Ni, 1.434 Å for Pd, 1.402 Å for Rh, 1.164 Å for Ru, 1.423 Å for Tc, and 1.270 Å for Zn, respectively. When the Wigner Seitz radius of transition metal X is larger than that of *β*-titanium, the volume change is greater than 0, and vice versa. As shown in Fig. 2, the relationship between Wigner Seitz radius and volume change can be fitted satisfactorily to linear function with an $R^2$ value of 0.9452,

$$y = 42.59x - 58.07 \tag{5}$$

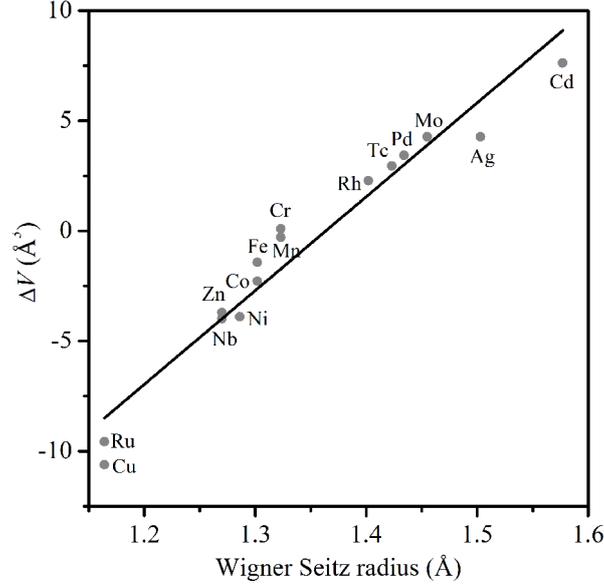

**Fig. 2.** Effect of Wigner Seitz radius ($r$) of dopants X on the volume change.

### Thermodynamic property of $Ti_{0.98}X_{0.02}$ system

In this section, the formation enthalpy of $Ti_{0.98}X_{0.02}$ system is calculated and listed in Table 2. We can find that the formation enthalpy values of $Ti_{0.98}Cr_{0.02}$, $Ti_{0.98}Mo_{0.02}$ and $Ti_{0.98}Nb_{0.02}$ are –9, –10 and –9 kJ/mol, respectively. These values close to the previously reported values of –9.77 kJ/mol for $Ti_{0.95}Cr_{0.05}$, –10.33 kJ/mol for $Ti_{0.95}Mo_{0.05}$ and –10.95 kJ/mol for $Ti_{0.95}Nb_{0.05}$ (first-principles calculation with PAW-GGA-PBE [4]). Thus, it can be concluded that the method of our formation enthalpy is reasonable.

**Table 2.** The DFT calculated lattice parameters ($a$, $b$, and $c$), equilibrium volume ($V_0$), volume change ($\Delta V$), zero-point energy ($\Delta ZPE$), formation enthalpy ($\Delta_f H$), and formation free energy ($\Delta_f F$) of $Ti_{0.98}X_{0.02}$ systems

| X | $a$, Å | $b$, Å | $c$, Å | $V_0$, Å³ | $\Delta V$, Å³ | $\Delta ZPE$, kJ/mol | $-\Delta_f H$, kJ/mol | $-\Delta_f F$, kJ/mol |
|---|---|---|---|---|---|---|---|---|
| Ag | 9.763 | 9.766 | 9.766 | 931.14 | 4.28 | 1.38 | 9 | 10 |
| Cd | 9.776 | 9.777 | 9.777 | 934.49 | 7.63 | 1.48 | 8 | 8 |
| Co | 9.742 | 9.742 | 9.742 | 924.58 | –2.28 | 1.38 | 9 | 9 |
| Cr | 9.751 | 9.750 | 9.750 | 926.95 | 0.10 | 1.44 | 9 | 9 |
| Cu | 9.712 | 9.712 | 9.714 | 916.25 | –10.61 | 1.38 | 9 | 10 |
| Fe | 9.745 | 9.745 | 9.745 | 925.43 | –1.43 | 1.30 | 9 | 9 |
| Mn | 9.749 | 9.749 | 9.749 | 926.57 | –0.29 | 1.44 | 10 | 11 |
| Mo | 9.755 | 9.770 | 9.770 | 931.14 | 4.28 | 1.45 | 9 | 9 |
| Nb | 9.736 | 9.736 | 9.736 | 922.88 | –3.99 | 1.46 | 10 | 11 |
| Ni | 9.731 | 9.739 | 9.739 | 922.97 | –3.89 | 1.34 | 9 | 10 |
| Pd | 9.762 | 9.762 | 9.762 | 930.29 | 3.43 | 1.36 | 10 | 10 |
| Rh | 9.756 | 9.759 | 9.759 | 929.14 | 2.28 | 1.24 | 11 | 11 |
| Ru | 9.717 | 9.716 | 9.716 | 917.29 | –9.57 | 2.75 | 11 | 12 |
| Tc | 9.759 | 9.761 | 9.761 | 929.81 | 2.95 | 2.42 | 10 | 10 |
| Zn | 9.737 | 9.737 | 9.737 | 923.16 | –3.70 | 1.83 | 9 | 9 |

As mentioned before, the formation enthalpy can be used to preliminarily judge the thermodynamic stability of materials [5,21,22,31]. From the Table 2, it can be found that the formation enthalpy of $Ti_{0.98}X_{0.02}$ system is between –8 and –11 kJ/mol, slightly than 0. The negative formation enthalpy indicates that formation of $Ti_{0.98}X_{0.02}$ from pure titanium and X is exothermic, namely, adding a small amount of transition metal X tends to enhance the thermodynamic stability of $β$-titanium at 0 K. In addition, since the more negative formation enthalpy, the stronger the enhancement ability, the enhancement ability of these metallic elements X on the $β$-titanium is in the order of Ru and Rh > Mn, Nb, Tc, and Pd > Ag, Co, Cr, Cu, Fe, Mo, Ni, and Zn > Cd, in which the elements Rh, Pd, and Ru are very expensive and the element Tc possesses radioactivity.

Eventually, the formation free energy of $Ti_{0.98}X_{0.02}$ system at 298 K is estimated to study the effect of temperature on stability of $β$-titanium, as shown in Table 2. It can be found that the formation free energy is more negative than formation enthalpy, suggesting that temperature play a positive role in the thermodynamic stability of $β$-titanium. This conclusion coincides with the fact that $β$-titanium is a stable phase at high temperature.

### *Electronic structure of $Ti_{0.98}X_{0.02}$ system*

In this section, the TDOS and PDOS of these systems are calculated to explore the strengthening mechanism of trace metallic element X on *β*-titanium from the electronic level, as shown in Fig. 3. Firstly, the TDOS curve of pure *β*-titanium is in good agreement with former report [6]. Thus, the simulation of TDOS in this paper is reasonable and can be used to research unknown system. Secondly, it is evident that the TDOS curves of $Ti_{0.98}X_{0.02}$ systems and *β*-titanium located in a range from –6 to 2.5 eV mainly originate from the contribution of Ti-*d* orbit. The results are in good agreement with Kuroda et al. proposed theory that the *d* electronic orbit plays an important role in mechanical properties and phase stability of *β*-titanium [32]. Thirdly, the Fermi level marked by a vertical dark dash line is set to zero energy in Fig. 3. The left side of the Fermi level is the valance band and the right side of the Fermi level is the conduction band. Obviously, we find that the value at the Fermi level of *β*-titanium is about 60 and the values at the Fermi level of $Ti_{0.98}X_{0.02}$ systems are in the range from 15 to 30. The value at Fermi level is related to the phase stability, the lower value at Fermi level, the more structure stable. So, the greatly reduced the value at Fermi level indicates that all dopants X in this study help to strengthen the structural stability of *β*-titanium at 0 K. This can be attributed that the transition metal elements X can provide additional *d*-type electrons required to improve stability of *β*-titanium. From these discussions, we explain that the addition of trace transition metal X plays a positive role in the stability of *β*-titanium from the atomic and electronic level.

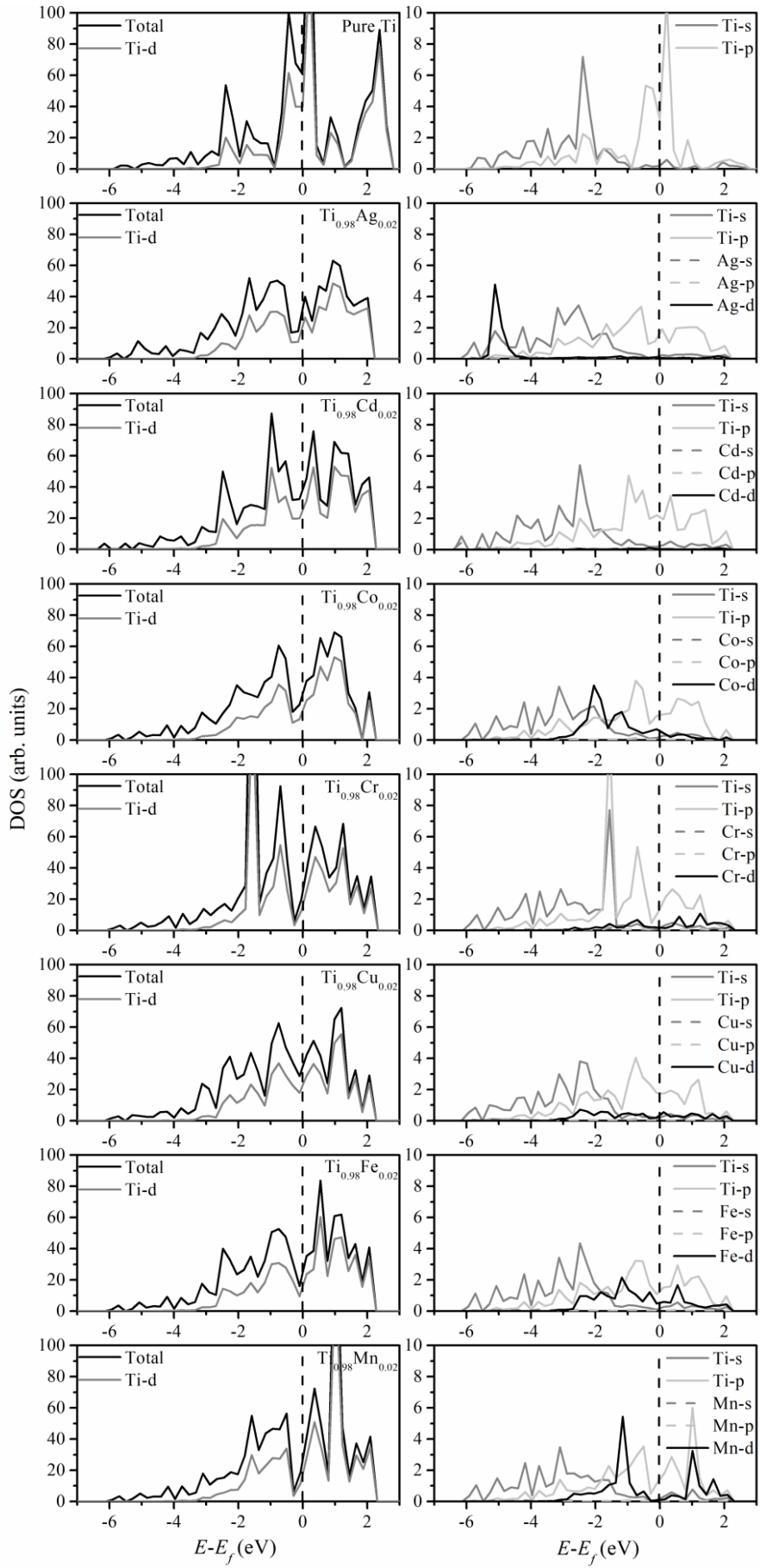

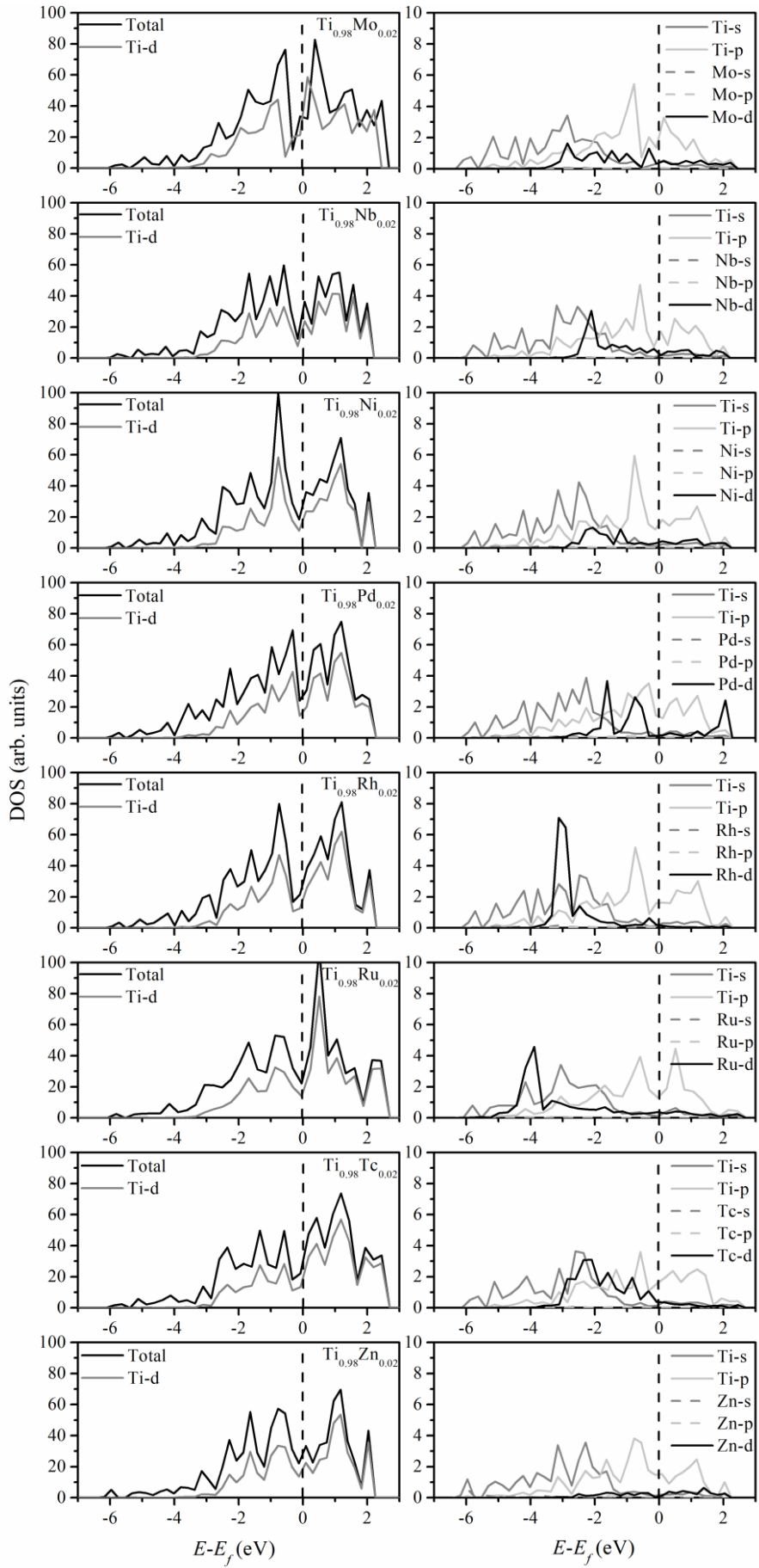

**Fig. 3.** The DFT calculated TDOS (left) and PDOS (right) of $Ti_{0.98}X_{0.02}$ systems and $\beta$-titanium. A vertical dotted line denotes the Fermi energy.

To understand the bonding characteristic more intuitively and vividly, the difference charge densities ($\Delta Q$) on the (1 0 0) plane of $Ti_{0.98}X_{0.02}$ systems are plotted by VESTA package and showed in Fig. 4. In each plot, the doped element atom is at the central position. The black regions ($\Delta Q > 0$) are electrons accumulating, while those in white ($\Delta Q < 0$) are electrons loss. From the Fig. 4, the bonding electron density between Ti and solute atoms are obviously increased and form the ring-type features, leading to an increase in bond strength and indicating an effective solid-solution strengthening the effect. Moreover, the region with a higher charge density suggests stronger bond formation. Since the black area means an increase in the electron density, the deeper the black, the stronger the bond between Ti and solute atoms. According to difference charge density plot, the bonding ability between Ti and platinum metals (Pd, Rh, Ru, and Tc) stronger than other transition metals within the scope of our study, which is consistent with the conclusion of formation enthalpy.

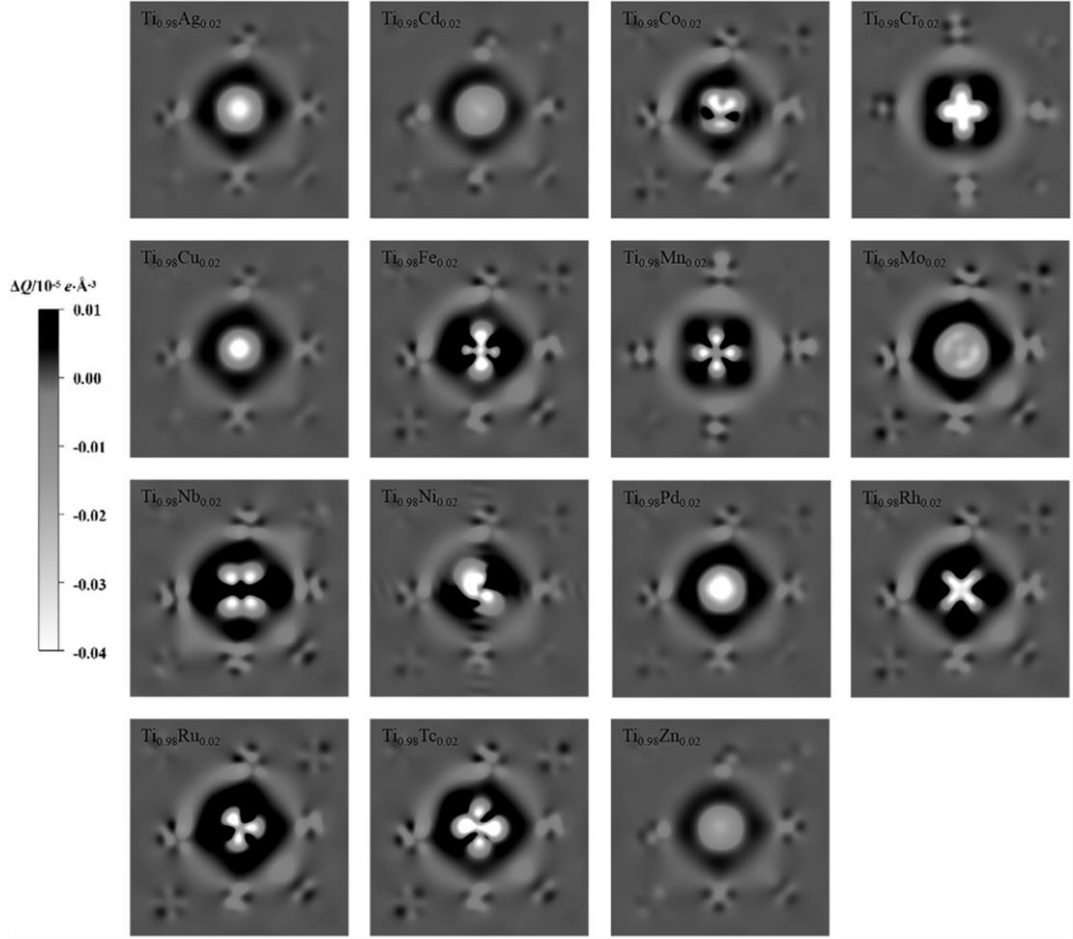

**Fig. 4.** The difference charge densities on the (1 0 0) plane of $Ti_{0.98}X_{0.02}$ systems. White and black color represent charge loss and charge accumulation, respectively.

*Elastic constants and mechanical stability*

Elastic constants can make us fully understand the stability of *β*-titanium doping with small amount of transition metal X. In this section, we estimate the elastic constants of $Ti_{0.98}X_{0.02}$ systems to investigate the mechanical stability of $Ti_{0.98}X_{0.02}$ systems in the ground state.

The elastic constants of pure *β*-titanium are directly calculated by the VASP code and listed in Table 3. Previous experimental results of *β*-titanium are also introduced in Table 3 for comparison to verify the accuracy of present calculations. From the Table 3, we find that the calculated $C_{11}$ and $C_{44}$ are 93.2 and 40.5 GPa agreeing well with experimental values at 97.7 and 37.5 GPa, but the theoretical and experimental values for $C_{12}$ is 114.0 and 82.7 GPa, respectively [27]. The large difference in $C_{12}$ results in a completely different theoretical and experimental tetragonal shear modulus $C'$. The theoretical tetragonal shear modulus is –10.4 GPa at 0 K, less than 0, while the experimental value is 7.5 GPa at 1273 K, greater than 0 [33]. As mentioned above, a negative tetragonal shear modulus means that cubic structure is an unstable, vice versa [6,27,28]. It is concluded that $β$-titanium is unstable at 0 K from the DFT simulating results but it is stable at 1273 K from the experimental values. This phenomenon is in accordance with a fact that $β$-titanium is a stable structure when the temperature exceeds 1155 K [34]. To sum up, the method of calculated elastic constants in this study is rational and can be applied reliably to predict the elastic constants of the system in which experimental measurements have not been done. Then, the same method is used to estimate the elastic constants of systems that have been selected in the previous section.

**Table 3.** The elastic coefficient ($C_{ij}$,GPa) and tetragonal shear modulus ($C'$,GPa) of $Ti_{0.98}X_{0.02}$, systems and $β$-titanium calculated from DFT simulations

| System | $C_{11}$ | $C_{22}$ | $C_{44}$ | $C'$ | Stability |
|---|---|---|---|---|---|
| $Ti_{0.98}Ag_{0.02}$ | 99.76 | 115.06 | 48.02 | -7.65 | No |
| $Ti_{0.98}Cd_{0.02}$ | 100.00 | 119.25 | 46.34 | -9.63 | No |
| $Ti_{0.98}Co_{0.02}$ | 96.77 | 116.00 | 41.89 | -9.62 | No |
| $Ti_{0.98}Cr_{0.02}$ | 95.71 | 115.61 | 38.87 | -9.95 | No |
| $Ti_{0.98}Cu_{0.02}$ | 104.09 | 113.78 | 40.69 | -4.85 | No |
| $Ti_{0.98}Fe_{0.02}$ | 99.13 | 119.00 | 37.94 | -9.94 | No |
| $Ti_{0.98}Mn_{0.02}$ | 100.49 | 114.53 | 38.90 | -7.02 | No |
| $Ti_{0.98}Mo_{0.02}$ | 103.26 | 108.58 | 44.46 | -2.66 | No |
| $Ti_{0.9583}Mo_{0.0417}$[a] | 110.35 | 112.44 | 42.45 | -1.05 | No |
| $Ti_{0.98}Nb_{0.02}$ | 101.28 | 118.39 | 47.40 | -8.56 | No |
| $Ti_{0.9375}Nb_{0.0625}$[a] | 109.73 | 115.23 | 43.88 | -2.75 | No |
| $Ti_{0.98}Ni_{0.02}$ | 94.87 | 109.21 | 45.05 | -7.17 | No |
| $Ti_{0.98}Pd_{0.02}$ | 103.46 | 107.59 | 43.46 | -2.07 | No |
| $Ti_{0.98}Rh_{0.02}$ | 102.59 | 106.43 | 45.80 | -1.92 | No |
| $Ti_{0.98}Ru_{0.02}$ | 104.82 | 108.98 | 38.50 | -2.08 | No |
| $Ti_{0.98}Tc_{0.02}$ | 102.03 | 106.61 | 36.87 | -2.29 | No |
| $Ti_{0.98}Zn_{0.02}$ | 105.49 | 113.82 | 39.76 | -4.17 | No |
| $\beta$-Ti | 93.20 | 114.00 | 40.50 | -10.40 | No |
| $\beta$-Ti$_{Exptl.}$ (1273 K)[b] | 97.70 | 82.70 | 37.50 | 7.50 | Yes |

[a] Yao et al., DFT calculation [35].

[b] Ledbetter et al., experimental measure [33].

The array of elastic constants for $Ti_{0.98}X_{0.02}$ systems are obtained by using DFT simulations and summarized in Table 3. Not only is the elastic constants of $Ti_{0.98}Nb_{0.02}$ consistent with previous reported value of 109.73 ($C_{11}$), 115.23 ($C_{12}$) and 43.88 GPa ($C_{44}$) evaluated from DFT simulating $Ti_{0.9375}Nb_{0.0625}$ system, but the elastic constants of $Ti_{0.98}Mo_{0.02}$ is also in good agreement with previously reported value of 110.35 ($C_{11}$), 112.44 ($C_{12}$) and 42.45 GPa ($C_{44}$) evaluated from DFT simulation of $Ti_{0.9583}Nb_{0.0417}$ system [35].

Owing to the elastic stability reflecting crystal structure stability to some extent, we further investigate the elastic stability of these systems using elastic constants. From Table 3, we first find that the elastic constants of these systems all dissatisfy the Eq. (4). It is concluded that these systems are unstable, which means the concentration of transition metal X in this study is not enough to make the $\beta$-titanium elastically stable at 0 K. What's more, the values of tetragonal shear modulus of these systems are all larger than $\beta$-titanium, implying that adding these elements is beneficial to the elastic stability of $\beta$-titanium at 0 K, which agree with the results of formation enthalpy. So, it can infer that if we continuously add transition metal to $\beta$-titanium, we can find a concentration of X that is able to make $\beta$-titanium elastically stable at 0 K.

## Conclusions

To improve our fundamental understanding of the effect of trace metallic elements on $β$-titanium, DFT simulations are proposed to systematically investigate the structure, formation enthalpy, formation free energy, elastic constants and electronic structure of $Ti_{0.98}X_{0.02}$ systems (X=Ag, Cd, Co, Cr, Cu, Fe, Mn, Mo, Nb, Ni, Pd, Rh, Ru, Tc, and Zn). We disclose a favorable linear function between the Wigner Seitz atomic radius and corresponding volumetric change. Furthermore, the negative formation enthalpy suggests that adding trace transition metal tends to improve thermodynamic stability of $β$-titanium in the ground state. This interesting behavior is well explained by the displacement of TDOS at the Fermi level.

The elastic constants of $Ti_{0.98}X_{0.02}$ systems are calculated to investigate the mechanical stability of all compounds in the ground state. The results show that adding trace transition metal can improve value of tetragonal shear modulus of $β$-titanium but can't make the tetragonal shear modulus greater than 0 GPa, which means the concentrations of these transition metals in our study are not enough to make the $β$-titanium elastically stable at 0 K.

To sum up, transition metal elements X have a positive effect on the stability of $β$-titanium, but the concentration of X is not enough to make the $β$-titanium stable at 0 K in this study. This work provides valuable insight into the nature of the important stability of $β$-titanium structures alloyed with diverse transition metal elements.

**Acknowledgments**

This work is supported by the project to strengthen industrial development at the grass-roots level (Project Number TC160A310/19), Young science and technology talents upgrading scheme (Project Number 18SG-15) and Technical innovation project of Shanghai research institute of materials (Project Number 19SG-04). Z. Gao acknowledges financial support from MOE tier 1 funding of NUS Faculty of Science, Singapore (Grant No. R-144-000-402-114).